\documentstyle[11pt,newpasp,twoside]{article}
\input psfig

\def\edcomment#1{\iffalse\marginpar{\raggedright\sl#1\/}\else\relax\fi}
\reversemarginpar
\def\ltsim{\hbox{\raise 2pt \hbox {$<$} \kern-1.1em \lower 4pt \hbox {$\sim$}}}
\def\ltapprox{\hbox{\raise 2pt \hbox {$<$} \kern-1.1em \lower 5pt \hbox 
{$\approx$}}}
\def\gtsim{\hbox{\raise 2pt \hbox {$>$} \kern-1.1em \lower 4pt \hbox {$\sim$}}}
\def\gtapprox{\hbox{\raise 2pt \hbox {$>$} \kern-1.1em \lower 5pt \hbox 
{$\approx$}}}

\def\arcsec{$^{\prime\prime}$}

\def\etal{{et al.~}}

\begin{document}
\title{Clusters of galaxies in radio}
 \author{Luigina Feretti}
\affil{Istituto di Radioastronomia CNR \\Via P. Gobetti 101, 40129 Bologna, 
Italy}

\begin{abstract}
Recent results on the radio emission from galaxy clusters
are reviewed, with emphasis both on the radio galaxies and on the
diffuse radio emission from the intracluster medium. 
We show that the formation of the tailed  morphology in cluster radio
galaxies is due to the effect of the  
cluster gaseous environment and of its
large scale motions in merging clusters.
From Chandra X-ray data, it has also become evident 
that the gas in the hot cluster atmospheres may be displaced by the
relativistic plasma.
The statistical characteristics of  radio galaxies
are briefly discussed, and are shown to be surprisingly similar
for sources both inside and outside  rich clusters.
The study of diffuse radio  sources originating from the
intracluster medium (halos and relics) has presently become 
a major area of investigation. 
Halos and relics are detected in massive X-ray luminous clusters
which are undergoing violent merger processes. The power of radio halos
is correlated with the cluster X-ray luminosity and mass.

\end{abstract}

\section{Introduction}

Studies at radio wavelengths allow the investigation
of an important aspect of  cluster formation and
evolution.
The radio emission in clusters of galaxies originates
mainly from  individual galaxies, which have
been imaged over the last decades with sensitive radio
telescopes. 
Powerful radio galaxies are the classic FR~I and FR~II sources
(with luminosity at 1.4 GHz lower and higher 
than $\sim$10$^{24.5}$ W Hz$^{-1}$, respectively,  Fanaroff-Riley 1974).
The radio emission  from these galaxies often extends well
beyond their optical boundaries, out to hundreds of
kiloparsec, so one might expect the intracluster medium (ICM) to 
affect their structure. This interaction is indeed observed
in extreme examples: the existence of radio galaxies
showing distorted structures (tailed
  radio sources), and 
radio sources filling X-ray cavities at the
center of some clusters.

Moreover, it is important to understand whether and how the cluster
environment plays any role in the statistical radio properties
of galaxies, i.e. their probability of forming radio
sources. 
Now, thanks to the high sensitivity of radio
observations, many low power radio
galaxies have been detected.
Deep radio surveys of clusters have been used to 
shed light on the connection between the cluster environment 
and the formation of radio galaxies, and to investigate
the starburst activity in cluster galaxies.

Another aspect of cluster radio emission is represented
by the  large-scale diffuse radio sources, which cannot
be obviously associated with the individual galaxies, thus
demonstrating the existence of non-thermal processes
in the ICM.
It is well established that a number of clusters exhibits
halos, relics and mini-halos. Their synchrotron radio emission 
implies magnetic fields
of the order of $\sim$0.1-1 $\mu$G, and 
relativistic particles of Lorentz factor $\gamma >>$1000 
and energy density of $\sim$10$^{-14}$-10$^{-13}$ erg cm$^{-3}$.
The importance of these sources is that they are large scale
 features, which are related to other cluster
properties in the optical and X-ray domain, and are
thus directly connected to the cluster history
and evolution.
Halos and relics are found in clusters which have recently undergone
a  merger event, thus leading to the idea that they 
originate from particle acceleration
in cluster merger turbulence and shocks.
The  formation and evolution of these sources
is however still under debate: radio emitting electrons could 
be reaccelerated cosmic rays, or accelerated  from the 
thermal population, or could be produced as a result
of the interaction between cosmic-ray protons and the ICM
(secondary electrons models).
We summarize the current knowledge
on these sources, from an observational point of view.

Throughout the paper, a Hubble constant
H$_0$ = 50 km s$^{-1}$ Mpc$^{-1}$
and a deceleration parameter q$_0$ = 0.5 are adopted.

\section{Cluster radio galaxies}

Recent results on the thermal gas in clusters
of galaxies has revealed  a significant
amount of spatial and temperature structure,
indicating that clusters are dynamically evolving
by accreting gas and galaxies and by merging with other
clusters/groups (roughly every few Gyrs).
Simulations suggest that the ICM within clusters is
violent, filled with shocks, high winds and turbulence.
This gas  can interact with a
radio source  in different ways:
modifying its morphology via ram-pressure, confining the
radio lobes, possibly feeding the active nucleus, affecting
the star formation.
We discuss below some of the recent results on these topics
(see also the review by Feretti \& Venturi 2002).

\subsection{Interaction between the radio galaxies and the ICM}

\subsubsection{Tailed radio galaxies.}

\begin{figure}
\centerline{\hbox{
\psfig{figure=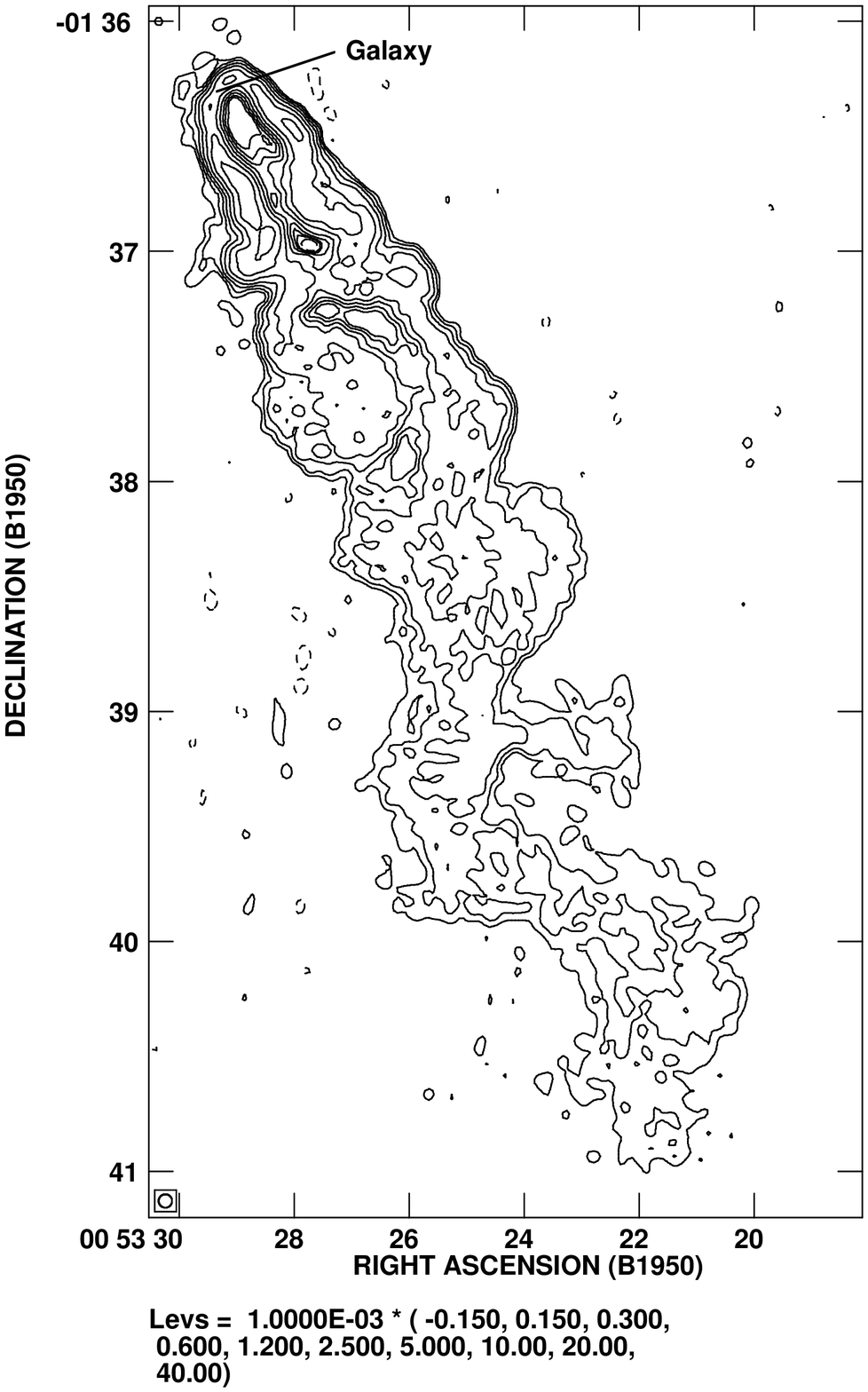,height=8cm}
\psfig{figure=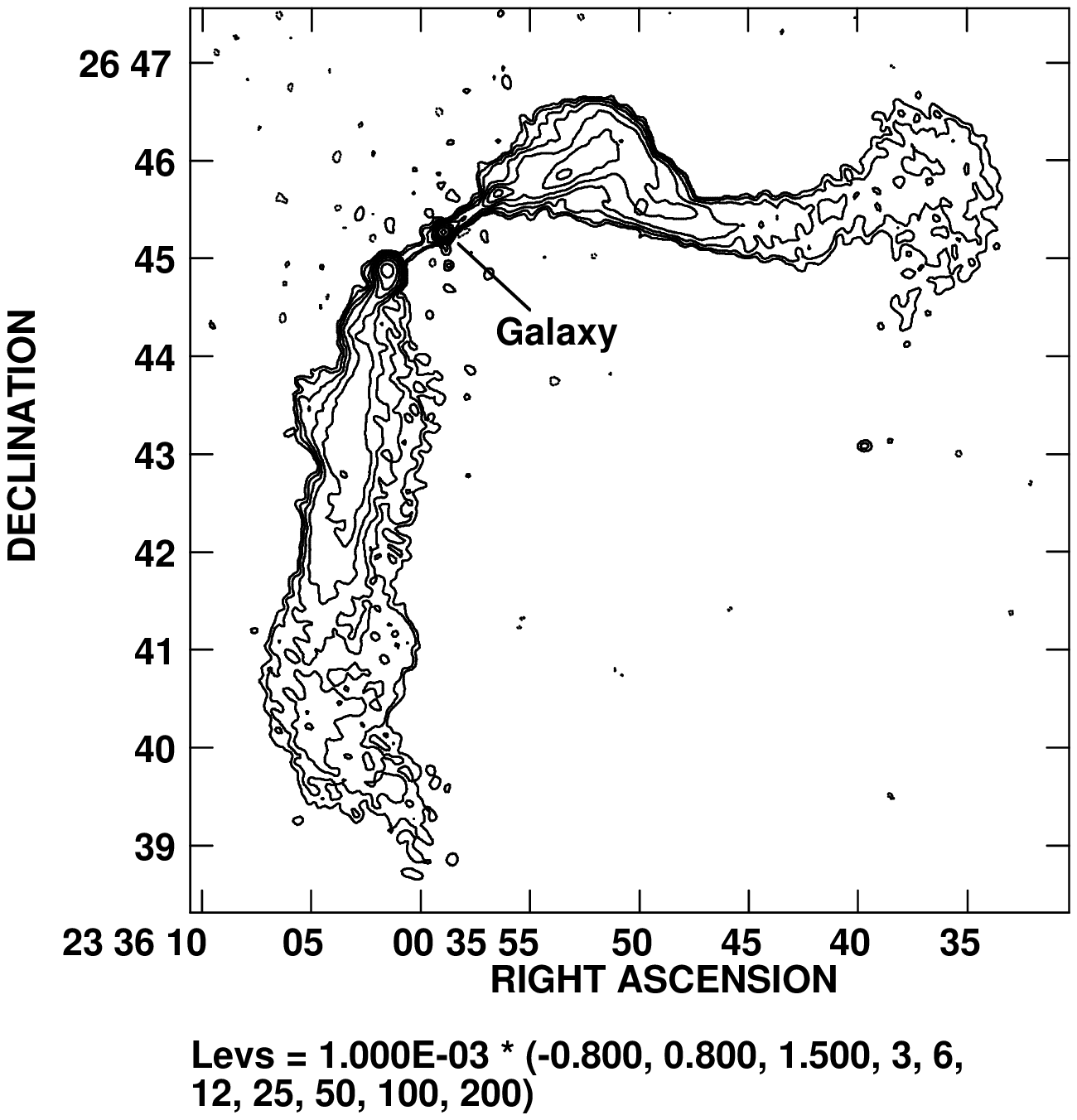,height=8cm}
}}
\caption{ Examples of tailed radio galaxies: 
the NAT  0053-016 in the cluster A119 (left panel) and 
the WAT 3C465 in the cluster  A2634 (right panel).
The location of the optical galaxy is indicated.}
\end{figure}

\begin{figure}
\centerline{\psfig{figure=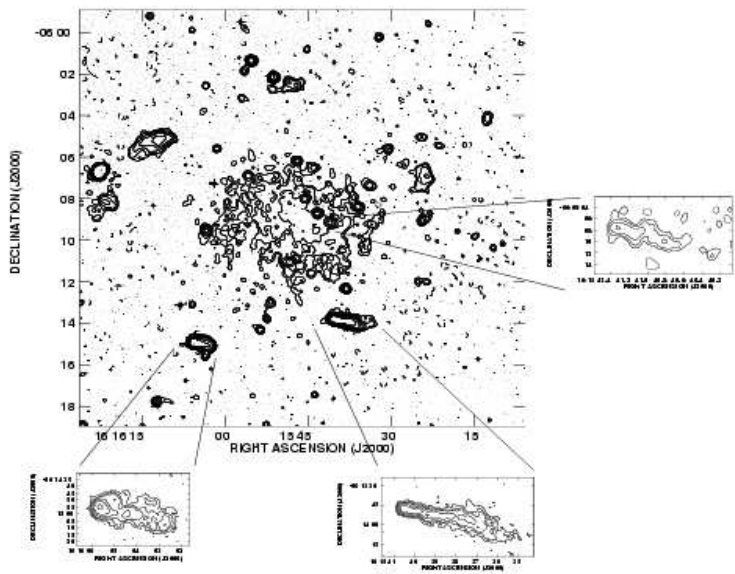,width=10cm,clip=}}
\caption{
Radio image of the cluster A2163 at 1.4 GHz, with angular resolution of 15\arcsec.
The structure of tailed radio galaxies as detected at higher resolution
is shown in the insets. The tails are oriented in the same
direction. }
\end{figure}

A dramatic example of the interaction 
of the radio galaxies with the ICM is represented
by the  tailed radio galaxies, i.e. FR~I sources where the
large scale low-brigthness emission is bent toward the same
direction, forming features similar to tails.
 These radio galaxies were originally distinguished in
two  classes: narrow-angle tailed sources (NAT), which  are "U" shaped
with a small angle between the tails,
and wide-angle tailed sources (WAT), which are "V" shaped with a 
larger angle between the tails (see Fig. 1). 
We note that distortions in FR~IIs are marginal and only present
in weak structures.

The standard interpretation of the tailed radio morphology is that the
jets are curved by ram  pressure from the high-velocity host
galaxy moving through the dense ICM, whereas
the low brightness tails are material left behind by the galaxy motion.
The ram pressure model, first developed  by Begelman
\etal (1979), can explain the radio jet deflection when the galaxy velocity 
with respect to the ICM is of the order of  $\sim$ 1000 km s$^{-1}$.
This model can be successfully applied to the NAT sources, whose parent
galaxies are located at any distance from the 
cluster center. 
Some clusters, however, contain tailed radio galaxies
with the tails oriented in the same direction (e.g., A2163, Fig. 2,
A119, Feretti \etal 1999, A520, Govoni \etal 2001b).
This can  hardly be explained by the above model, since it is
unlikely that the parent galaxies are moving in the same direction.

Also the interpretation  of WAT sources is quite 
problematic, since these sources are generally associated with 
dominant cluster galaxies moving very slowly ($<$ 100 km s$^{-1}$) relative to
the cluster velocity centroid.
Such slow motion is insufficient to bend the jets/tails of
WATs to their observed curvature by ram pressure.
It has then been suggested that WATs must be shaped, at least in part, 
by other ram pressure gradients not arising from motion of the host galaxy, but
produced by  mergers between clusters 
(Loken \etal 1995,  G\'omez \etal 1997).
Numerical simulations lead support to this idea: peak  
gas velocities well in excess of 1000 km s$^{-1}$ at various stages 
of the cluster merger evolution are found, which generally do not
decay below 1000 km s$^{-1}$ for nearly 2 Gyr after the core
passage. This is consistent with the observations, as
modeled  in the cluster A562 (Fig. 3).

The cluster merging events could play an important role also 
in the case of  NAT sources. 
The similar orientation
of tailed radio galaxies mentioned above
favours the idea that  the dominant effect in the
formation of the tailed
morphology is merger-induced bulk gas motion, at least in some
clusters (Bliton \etal 1998).

\begin{figure}
\centerline{\psfig{figure=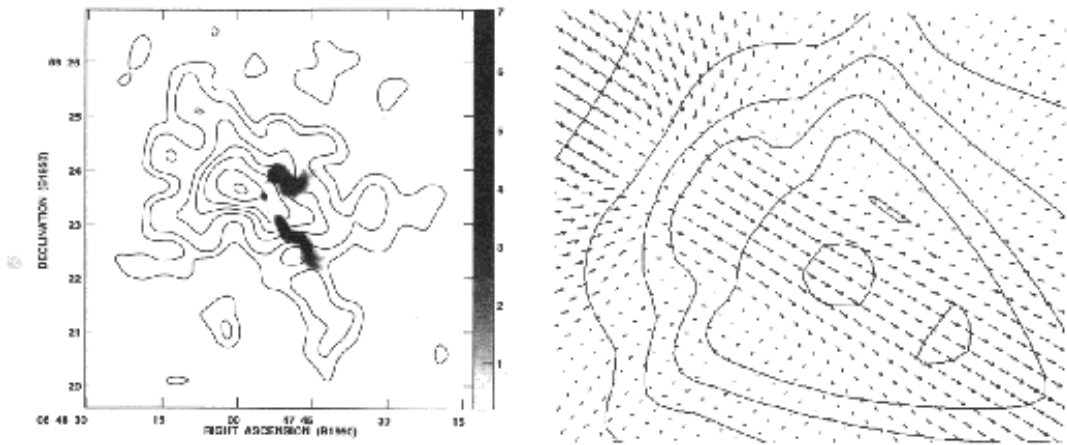,height=6cm,clip=}}
\caption{Left panel: overlay of a 6 cm radio map grey-scale
onto a ROSAT X-ray surface brightness contour  map of A562. 
Right panel: Overlay of a synthetic X-ray image of a cluster merger
onto a velocity vector field that represents the gas velocity. Note that
the X-ray contours in the left panel look very similar to the synthetic
X-ray image and that the radio tails are in the direction of 
the gas velocity (from G\'omez et al. 1997).}
\end{figure}

\subsubsection{Radio emission in X-ray cavities.}

A clear example of the interaction
between the radio plasma and the  hot intracluster medium was found 
in the Rosat image of the Perseus cluster (B\"{o}hringer \etal 1993), 
where X-ray cavities associated with the inner radio lobes
to the north and south of the bright central 
radio galaxy 3C84 have been first detected.
The high spatial resolution of the Chandra X-ray Observatory
has confirmed the presence of such X-ray holes
(Fabian \etal 2000),  coinciding with the radio lobes
and showing rims  cooler than the surrounding gas.
Chandra has  permitted the detection of X-ray deficient bubbles 
in the inner region of many cooling flow clusters, e.g. 
Hydra A, A2052, A496, A2199.
Another example is the cluster RBS797 (Schindler \etal 2001), at z=0.35,  
 which reveals two pronounced symmetric X-ray minima, approximately in the 
E-W direction.
Preliminary  low resolution radio data
at 8.3 GHz show the presence of a radio source elongated 
in the same direction
as the depressions observed in  X-rays (Fig. 4, see also
De Filippis, Schindler \& Castillo-Morales 2002). 
A study with higher resolution radio data is in progress.

\begin{figure}
\centerline{\psfig{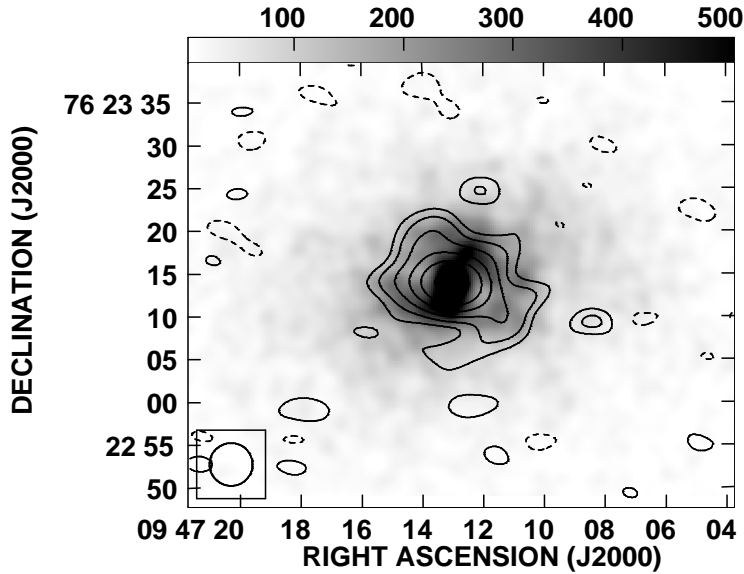}}
\caption{Radio contours superimposed on the smoothed 
Chandra X-ray 
grey-scale image of the core of RBS797.
The radio source shows an extension in the same direction
of the cavities  observed in the X-rays.}
\end{figure}

The interpretation of the X-ray cavities is that the
X-ray material which cooled at the cluster center is
pushed away by the radio source, and 
the bright rim detected in X-ray is due to
a mild shock driven by the expanding radio lobe
(Soker, Blanton \& Sarazin 2002, Nulsen \etal 2002).
X-ray depressions coincident with the position of the radio jets
were also found from Rosat data 
in 3C465 (Schindler \& Almudena Prieto 1997), at the center of the 
non cooling flow cluster A2634. Also in this case, the
inhomogeneous gas distribution could be 
produced by the displacement of the gas due to the radio plasma
flow.

All these examples revealed a wealth of new information
on the interaction between radio 
lobes and thermal gas (see Forman \etal 2002 for a review).
In addition, it has been suggested 
that the jets of radio sources could provide the energy required to stop
copious amounts of gas from cooling to low temperatures in the cooling
cores of clusters  (B\"ohringer \etal 2002).

\subsection{Trigger of radio emission}

An important issue is to understand whether and how the cluster
environment plays any role in the statistical radio properties
of galaxies, in particular their probability of forming radio
sources.
The high density of galaxies within clusters, especially in the
innermost cluster regions, and the peculiar velocities of
galaxies, most extreme in merging clusters, enhance the probability of
galaxy-galaxy interactions. These special conditions raise the
question whether cluster galaxies have enhanced probability
of developing a radio source, or whether they tend to have more
powerful and long lived radio emission. This subject has been 
first investigated by Fanti (1984), and further by 
Ledlow \& Owen (1996).
The most striking result 
is that statistical properties of  radio galaxies
are surprisingly similar
for sources both inside and outside rich clusters.
Even for cluster galaxies the only relevant parameter
seems to be the optical magnitude, i.e. brighter galaxies have a higher 
probability of developing a radio galaxy. 
Furthermore, the radio luminosity function (RLF)
 is indipendent on richness class, Bautz-Morgan
or Rood-Sastry cluster class.

The universality of the local RLF for early type galaxies can be 
applied also to merging clusters. Apparently, the enhanced 
probability of galaxy interaction in merging clusters has no effect
on the  probability of galaxies to develop a radio active galactic nucleus
 in their centres. The central clusters of
the Shapley Concentration, the largest concentration of merging
clusters in the local Universe,  were observed at radio wavelengths, and
the results show that 
even this exceptionally unrelaxed environment does not seem not to increase
the probability of elliptical galaxies to develop a radio source
(Venturi \etal 2001).

The large scale environment, and in particular
cluster merging processes, could  influence  the starburst emission.
From numerical simulations, Evrard (1991)
predicts an enhanced star formation  
due to the increasing external pressure as the galaxy 
falls into the dense intracluster 
medium. Different results are obtained by the modeling
of Fujita \etal (1999), who
argue that gas stripping during cluster collision is the dominant
factor. This leads to a decrease in  star formation rate, hence
weakening the starburst phenomenon during the merger.

At low radio 
power (LogP$_{1.4 GHz}$(W Hz$^{-1}$) \ltsim~22), star formation
becomes the dominant mechanism in the production of radio emission
(Dwarakanath \& Owen 1999), therefore deep radio surveys
of distant clusters and of dynamically unrelaxed clusters may
shed light on the connection between cluster merger and  triggering
of starburst activity. 
A limited number of observational studies has been conducted so far, 
with the  results presently  still controversial
(Owen \etal 1999; Baldi, Bardelli \& Zucca 2001; Miller \& Owen 2001).

\section{Diffuse radio emission from the intracluster medium}\label{sec:im}

A number of clusters of galaxies is known to contain large-scale
diffuse radio sources 
 which have no obvious connection with the cluster
population of galaxies, but are rather associated with the intergalactic
medium. Thus, these sources 
demonstrate the existence of high energy relativistic
electrons and magnetic fields in the ICM (see Giovannini \& Feretti 2002
for a review).

\begin{figure}
\centerline{\psfig{figure=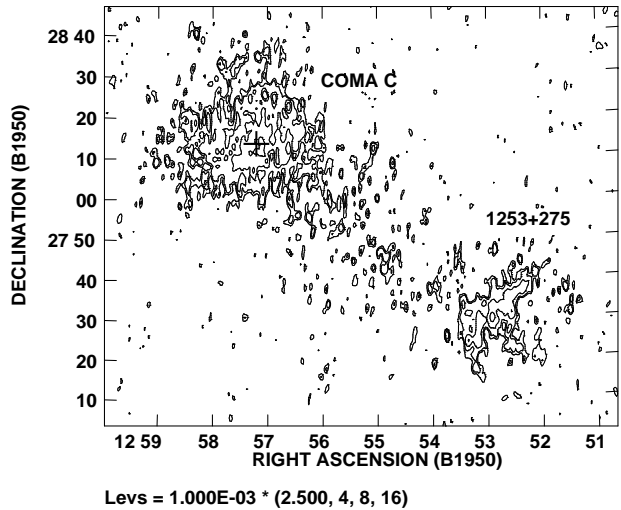,width=10cm,angle=270}}
\caption{
Radio map of Coma at 90 cm,
with angular resolution of 55\arcsec$\times$
125\arcsec(RA $\times$ DEC) after subtraction of individual
discrete sources. A cross indicates the
cluster center. The radio halo Coma C and
the relic 1253+275 are visible. 
}
\end{figure}

Radio halos are extended diffuse radio sources
with low surface brightness
($\sim$1 $\mu$Jy arcsec$^{-2}$ at 1.4 GHz), permeating the cluster center.
They show an approximately regular shape, a steep radio spectrum
typical of aged radio sources ($\alpha > $ 1), and 
little or no polarized emission.
Halos are generally extended \gtsim~1 Mpc, but 
some smaller halos, down to \ltsim~500 kpc in size, have also 
been detected.
The prototype halo is Coma C in the Coma cluster (Fig. 5), 
first  classified by Willson (1970).

Relic sources are similar to halos in their low
surface brightness, large size and steep spectrum, but 
are located in  cluster peripheral regions. In most cases 
they show an elongated structure and are highly polarized.
Originally relic sources were suggested to be relics of currently non 
active galaxies. However,
no evidence has been  found to support this interpetation.
The prototype of the class is 1253+275,
in the Coma cluster (see Fig. 5), 
first classified by  Ballarati \etal (1981).
A spectacular example of two almost symmetric relics in the same
cluster is found in A3667 (R\"ottgering et al. 1997).

Feretti (2000) argued that halos and relics are not the same objects 
seen in projection, i.e. halos are really at the
 cluster center and not simply projected onto it.
In fact halos and relics may  have different
physical origins.

Mini-halos are diffuse extended radio sources
of moderate size ($\sim$ 500 kpc)
surrounding a  dominant powerful radio galaxy at
the  center of cooling flow clusters.
Examples of this class are detected in the Perseus  
and Virgo clusters. Gitti, Brunetti \& Setti (2002) argued that
a connection between the central radio galaxy 
and the mini-halo in terms of particle diffusion or
buoyancy is not possible in the Perseus mini-halo, and suggested that 
the relativistic electrons 
are reaccelerated by turbulence in the cooling flow 
region of the ICM.
Deeper studies of a large sample of mini-halos are needed to
clarify the role of the central radio galaxy 
for the injection of the electrons in  mini-halos.

Due to synchrotron and inverse Compton losses, the typical ageing 
time-scale of the relativistic electrons in the ICM is relatively
short ($\sim$10$^8$ yr, Sarazin  1999).
The difficulty in explaining radio halos and relics arises from the
combination of their large size and the short synchrotron
lifetime of relativistic electrons. 
The expected diffusion velocity of the electron population  is of the order
of the Alfv\'en speed ($\sim$ 100 km s$^{-1}$) making it difficult for the
electrons to diffuse over a megaparsec-scale region within their
radiative lifetime.
Thus the relativistic particles need to be reaccelerated by some mechanism, 
acting with an efficiency comparable to the energy loss
processes (Petrosian 2001).
We will show in the following that cluster mergers are likely
to supply energy to the halos and relics. 
In the relics, diffusive shock acceleration or adiabatic 
compression of fossil radio plasma by merger shock waves 
have been suggested (En{\ss}lin \& Br\"uggen 2002 and 
references therein).

\subsection{Recent studies}

Great attention has been devoted in recent years to the
study of cluster large-scale diffuse radio sources.
New halo and relic candidates were found from searches in the
NRAO VLA Sky Survey (Giovannini, Tordi \& Feretti 1999), 
in the Westerbork Northern Sky Survey (Kempner \& Sarazin 2001) 
and in the 
Sidney University Molonglo Sky Survey (Hunstead \etal 1999).
In addition,  detailed studies of individual halos and relics
have  been  performed, providing more information on
these sources.
High radio luminosity halos have been studied in distant clusters,
as A665 (z = 0.1818, Giovannini \& Feretti 2000),
A2163 (z = 0.203, Feretti \etal 2001), A2744 
(z = 0.308; Govoni \etal 2001b), and Cl 0016+16
(z = 0.5545; Giovannini \& Feretti 2000).
A powerful radio halo was found in the hottest known cluster
of galaxies (1E$0657-56$, kT = 15.6 keV; z = 0.296) by Liang \etal (2000).
From low frequency VLA observations, the existence of a halo
and a possible relic has been confirmed
in A754 (Kassim \etal 2001), where the presence of diffuse emission
was debated in the literature.

\subsection{Connection to cluster merger processes}

Unlike the presence of thermal X-ray emission, the
diffuse radio emission is not a common property 
in clusters of galaxies.
Giovannini Feretti \& Govoni (2000) report that, in 
a complete cluster sample, 5\% of clusters have a radio halo 
source and 6\% a peripheral relic source. 
The detection rate of diffuse radio sources
increases with the cluster X-ray luminosity, reaching $\sim$35\% in
clusters with X-ray luminosity 
larger than 10$^{45}$ erg s$^{-1}$ (in the Rosat band).

A possible explanation of the relative rarity of diffuse sources
is that they  are forming only in clusters which have undergone
recent major merger events. Indeed, the cluster merging would
supply the energy needed to reaccelerate the radiating 
particles all around  cluster. This condition is particularly
relevant  for the largest radio halos, given the
difficulty for radiating electrons to travel cluster-size distances
over their radiative lifetime.  

Several evidences point to the suggestion  that  
clusters with halos and relics are characterized by merging processes.
These clusters indeed show: 
\par\noindent
- substructures and distortions in the X-ray
brightness distribution (Schuecker \etal 2001), which can be interpreted as
the result of subclump interaction;
\par\noindent
- temperature gradients and  gas shocks, as found in
earlier temperature  maps with ASCA (e.g., Markevitch \etal 1998)
and recently with higher sensitivity and resolution with Chandra 
(Markevitch \& Vikhlinin 2001);
\par\noindent
-  large values of the dipole power ratio (Buote 2001), which indicates
high dynamical disturbance and departure from a virialized status;
\par\noindent
- more isolated position 
(Schuecker, this conference), giving additional
support to the idea that recent merger events 
 lead to a depletion of the nearest neighbours;
\par\noindent
- absence of a strong cooling flow
(e.g., Feretti 1999). This is further indication that a cluster has undergone 
a recent merger, as 
a strong merger  is expected to disrupt a cooling flow
(G\'omez \etal 2002); 
\par\noindent
- core radii significantly larger 
($>$99\% level using a KS test)
than those of clusters classified as single/primary
(Feretti 2000). Large core radii are usually 
expected in multiple systems 
in the process of merging (Jones \& Forman 1999);
\par\noindent
-  values of spectroscopic $\beta$  on average larger 
than 1 (Feretti 2000), indicating  the presence of substructure 
(Edge \& Stewart, 1991).

\begin{figure}
\centerline{\psfig{figure=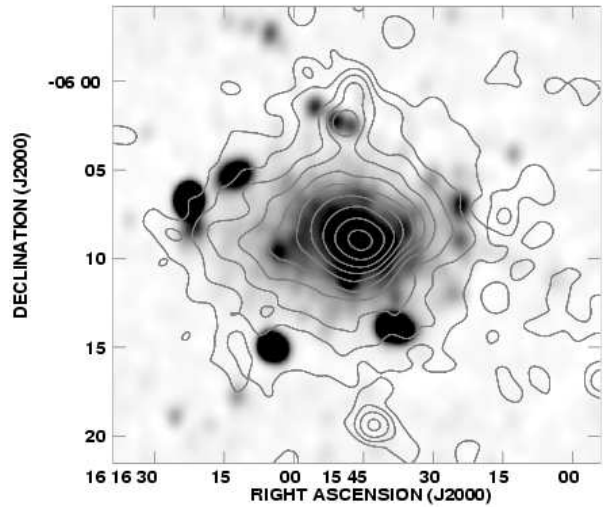,height=8cm}}
\caption{
Overlay of the X-ray contours onto the grey-scale radio map 
of A2163. The halo radio 
brightness  B$_{\rm radio}$ and the cluster 
X-ray brightness  B$_{\rm X-ray}$ are related  as
$ B_{\rm radio} \propto B_{\rm X-ray}^{0.64 \pm 0.05}$
(Feretti \etal 2001).
}
\end{figure}

In conclusion, there seems to be convincing evidence that 
diffuse sources are preferentially associated with high X-ray luminosity
clusters with mergers. Although  not all the  
merging clusters host a diffuse source, there is still 
no evidence of any radio halo in a cluster
where the presence of a merger has been clearly excluded.

\subsection{Connection between non-thermal and thermal emission} 

A morphological similarity between  the thermal gas distribution and the 
structure of the radio halos on the large scale 
has been  derived for clusters with good quality X-ray and radio data
(Govoni \etal 2001a, Feretti \etal 2001, see Fig. 6). 
A correlation between  the radio and X-ray brightness, 
$ B_{\rm radio} \propto B_{\rm X-ray}^b$,
has been found with  the parameter $b$  between 0.65 and 1. This indicates
that the energy density in the relativistic electrons 
is linked to the thermal energy density.
High resolution X-ray Chandra data 
have been recently obtained for the halo clusters
A665, A2163 (Markevitch \& Vikhlinin 2001),
1E 0657-56 (Markevitch \etal 2002) and 
A520 (Markevitch, this conference).
Preliminary data are also available for A2255
(Mazzotta \etal in preparation).
In A520,  a clear correlation
between the gas temperature and the radio brightness
seems to be present. In the other clusters,  temperature
gradients and gas shocks, which could
be related to the halos, are detected, 
 but the structures are more complicated.
Therefore, it is clear that the thermal and relativistic
plasma are connected, although the details need further
investigation. The radio - X-ray connection 
still points to a link between halos and cluster mergers.

\begin{figure}
\centerline{\psfig{figure=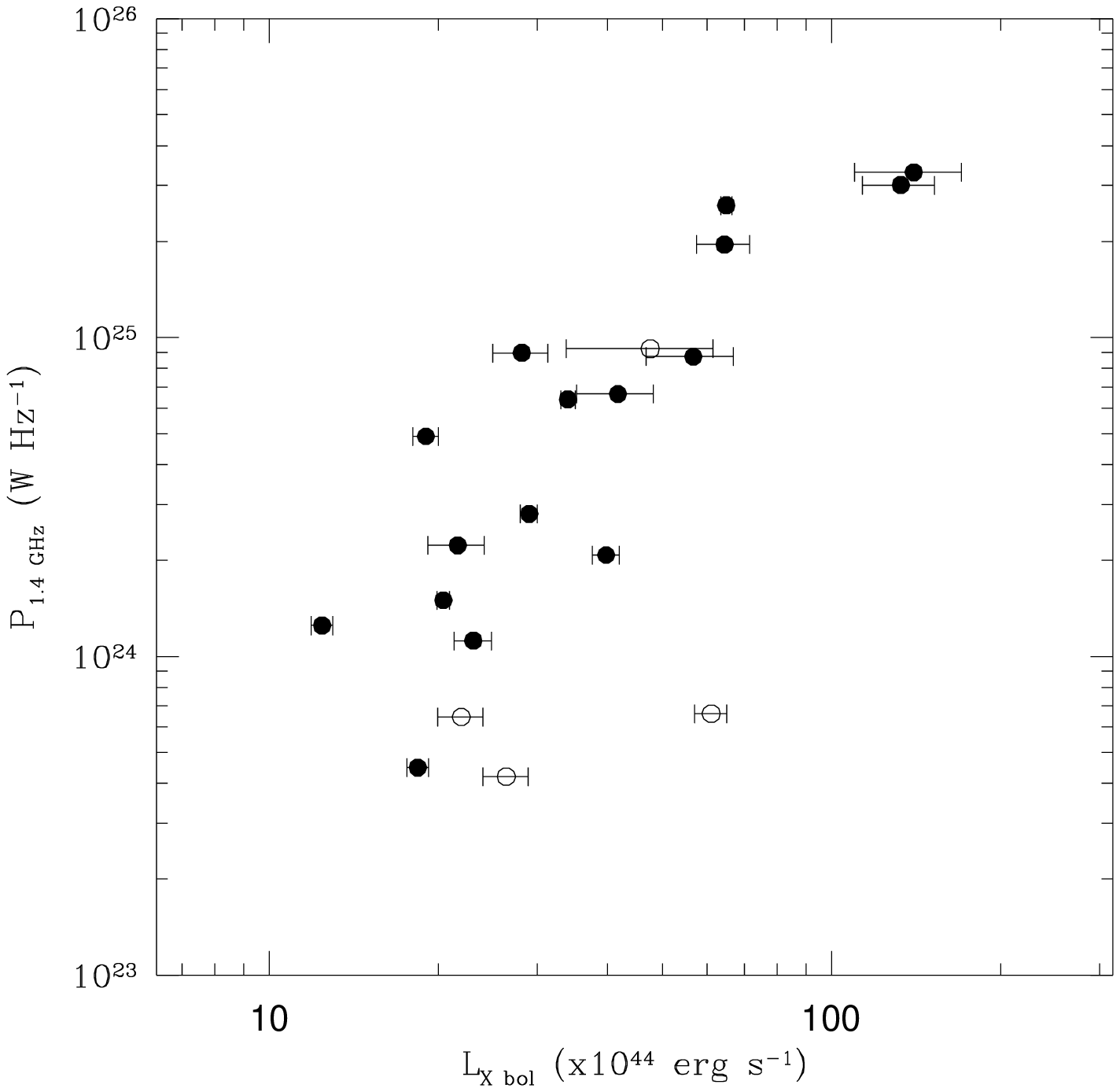,width=9cm}}
\caption{Halo monochromatic radio power at 
1.4 GHz versus the bolometric X-ray luminosity for sizes
$>$1 Mpc ($\bullet$) and $<$ 1 Mpc ($\circ$).
}
\end{figure}

\begin{figure}
\centerline{\psfig{figure=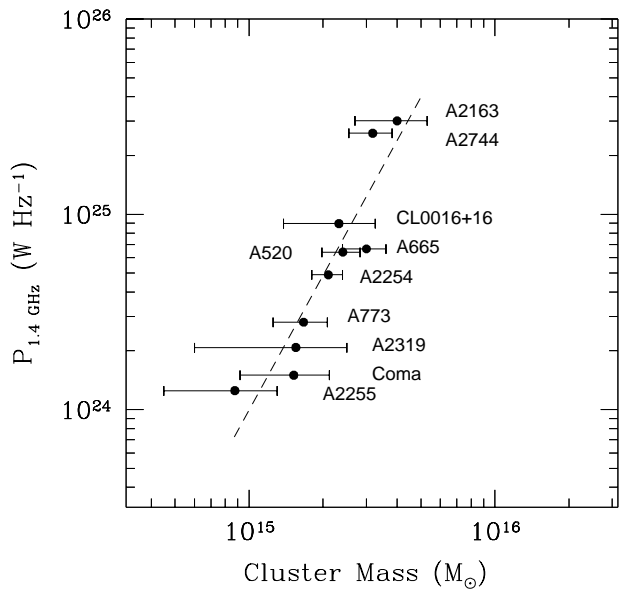,width=9cm}}
\caption{
Halo monochromatic radio power at 
1.4 GHz versus  total gravitational mass 
within 3 Mpc radius.
All the data are from  ROSAT, except A2163 from 
XMM (Monique Arnaud, private communication).
The best fit line is P$_{1.4 GHz}$ $\propto$ M$^{2.3}$.}
\end{figure}

\subsection{Correlation between radio and X-ray parameters} 

The presence of a correlation
between the halo radio power and the cluster X-ray luminosity
has been first noted by 
Liang \etal (2000) and confirmed by Feretti (2000). 
This correlation is presented in  Fig. 7, where 
solid circles refer to halos larger than 1 Mpc and open circles 
refer to  halos of smaller size. 
We stress here that 
the correlation applies to clusters showing major merger
events, and therefore  cannot be generalized to all
clusters. 
Among the clusters with high X-ray luminosity
and no radio halo, there are A478, A576, A2204,
A1795, A2029, all well known relaxed clusters with a massive
cooling flow.

Using only clusters with giant radio halos (solid circles), the best fit
between radio and X-ray luminosity is P$_{1.4 GHz} \propto L_{X}^{f}$, with 
$f$ in the range 1.6-2.0 (taking into account
values from different regression lines).
The addition of the smaller size halos does not change
 the overall slope of the  correlation, but adds more
dispersion to the X-ray data in the
region of lower radio power. Consequently,   
 at P$_{1.4 GHz}$ lower than $\sim$ 
3 10$^{24}$  W Hz$^{-1}$, the correlation seems to be virtually absent,
with the only indication that halos are present in clusters
with X-ray luminosity $>$10$^{45}$  erg s$^{-1}$.

An extrapolation of the large
radio halo correlation  to low radio and X-ray luminosities indicates 
that clusters with  L$_{X}$
 \ltsim~10$^{45}$ erg s$^{-1}$
would host halos of power of a few 10$^{23}$ W Hz$^{-1}$.
With a typical size of 1 Mpc, they would have a radio surface
brightness lower than current limits obtained in the literature
and  in the NRAO VLA Sky Survey.
Therefore, it is possible that 
future new generation instruments (LOFAR, SKA)
will allow the detection of low brightness/low power large halos 
in virtually all the merging clusters.
On the other hand, future highly sensitive 
data will clarify whether the correlation 
holds also at low power and for small size halos, or 
the cutoff in the X-ray luminosity suggested by Fig. 7 
is real. 

Since  cluster X-ray luminosity and mass are correlated
(Reiprich \& B\"ohringer 2002), the above correlation 
between radio power and X-ray luminosity could reflect a dependence
of radio power on cluster mass. 
Govoni et al. (2001b)  first obtained
this correlation using  6 halo clusters with a
homogeneously estimated gravitational mass M.
The plot shown in Fig. 8 includes  some more clusters
than in Govoni \etal (2001b). The best fit is 
P$_{1.4 GHz}$ $\propto$ M$^{2.3}$, which is
similar to what expected from simple theoretical considerations.
Actually, assuming that roughly  the energy
released in a merger shock is 
proportional to the gas density $\rho$ and to 
the subcluster velocity $v^3$, and that 
$\rho \propto$ M, and $v \propto {\rm M}^{1/2}$,
it is obtained that \.E $\propto$ M$^{5/2}$
 (see also Kempner \& Sarazin 2001).
A correlation P$_{1.4 GHz} \propto {\rm M}^{7/3}$
is predicted by Waxman \& Loeb (2000) in the case of 
shock-accelerated radio emitting electrons.

The particle reacceleration process may be more complicated,
involving also the merger turbulence (Brunetti \etal 2001).
Nevertheless, the indication from the above arguments is
that the  cluster mass seems to be a crucial parameter in the formation
of radio halos.  Since it likely that massive clusters undergo
several major mergers during their lifetime, 
we conclude that both past mergers and current mergers
are the necessary ingredients for the formation
and evolution of radio halos.

\section {Conclusions}

We have shown that the radio properties of clusters
of galaxies are strictly connected to the cluster formation
history and to the present cluster evolution.
The interaction between the radio galaxies and the 
hot cluster gas can now be studied with X-ray data
at high resolution, revealing unexpected phenomena.
The presence of non-thermal components in the intracluster
medium, i.e. relativistic particles and magnetic fields, is
now well established and is found to be related to the
existence of past and present merger events in  clusters
of galaxies. 

\acknowledgments

I would like to thank  Chorng-Yuan Hwang and Stuart Bowyer,
 and the local organizing committee for
organizing such an enjoyable and useful conference.
I am grateful to my collaborators G. Brunetti, I.M Gioia,
G. Giovannini, F. Govoni
 and T. Venturi for their contribution to this review.  
This work was partially funded by the Italian Space Agency.

\end{document}